\documentstyle[epsf,mlpp]{article}
\lefthead{LANDGRAF ET AL.}
\righthead{MASS DISTRIBUTION OF INTERSTELLAR DUST}
\authoraddr{M. Landgraf, 
NASA/Johnson Space Center, 
Mailcode: SN2,
2101 NASA Road 1,
Houston, TX 77058.
(e-mail: Markus.K.Landgraf1@jsc.nasa.gov)}
\slugcomment{To appear in the {\it Journal of Geophysical
Research} (JGR special section: ``Interstellar Dust and the
Heliosphere''), invited contribution}
\begin{document}
\title{Aspects of the Mass Distribution of Interstellar Dust
Grains in the Solar System from In-Situ Measurements}
\author{M. Landgraf}
\affil{NASA/Johnson Space Center, Houston, Texas, U.S.A.}
\author{W. J. Baggaley}
\affil{University of Canterbury, Christchurch, New Zealand}
\author{E. Gr\"un}
\author{H. Kr\"uger}
\author{G. Linkert}
\affil{Max-Planck-Institut f\"ur Kernphysik, Heidelberg, Germany}
\begin{abstract}
The in-situ detection of interstellar dust grains in the Solar System
by the dust instruments on-board the Ulysses and Galileo spacecraft as
well as the recent measurements of hyperbolic radar meteors give
information on the properties of the interstellar solid particle
population in the solar vicinity. Especially the distribution of grain
masses is indicative of growth and destruction mechanisms that govern
the grain evolution in the interstellar medium. The mass of an
impacting dust grain is derived from its impact velocity and the
amount of plasma generated by the impact. Because the initial velocity
and the dynamics of interstellar particles in the Solar System are
well known, we use an approximated theoretical instead of the measured
impact velocity to derive the mass of interstellar grains from the
Ulysses and Galileo in-situ data. The revised mass distributions are
steeper and thus contain less large grains than the ones that use
measured impact velocities, but large grains still contribute
significantly to the overall mass of the detected grains. The flux of
interstellar grains with masses $> 10^{-14}\;{\rm kg}$ is determined
to be $1\cdot 10^{-6}\;{\rm m}^{-2}\;{\rm s}^{-1}$. The comparison of
radar data with the extrapolation of the Ulysses and Galileo mass
distribution indicates that the very large ($m > 10^{-10}\;{\rm kg}$)
hyperbolic meteoroids detected by the radar are not kinematically
related to the interstellar dust population detected by the
spacecraft.
\end{abstract}
\begin{article}
\section{Introduction}
The measurement of interstellar dust grains by the dust detectors
on-board the Ulysses and Galileo spacecraft, allows us to determine the
grain mass distribution of the local interstellar dust
component. Impacts by interstellar dust grains were clearly identified
first in the Ulysses data after Ulysses' fly-by of Jupiter in February
1992 \cite{gruen93}. It was shown \cite{gruen94} that they can be
distinguished from interplanetary grains by their retrograde impact
direction which also coincides with the direction from which
interstellar neutral Helium atoms enter the Solar System
\cite{baguhl95a}. A further indication that Ulysses did detect
interstellar grains was the nearly constant rate and direction of impacts
on the Ulysses detector after the spacecraft left the ecliptic plane
and performed measurements at high latitudes
\cite{landgraf96}. Additionally, the measured impact velocities
measured by Ulysses after Jupiter fly-by indicated, although subject
to substantial uncertainties, that the dust velocities exceeded
the local escape velocity, even if radiation pressure effects were
neglected. The detection of interstellar dust grains by Ulysses were
confirmed by measurements with the Galileo dust detector
\cite{baguhl95a}.

The mass distribution of dust grains in the galactic interstellar
medium is an indicator for the grain growth and destruction processes
inside the medium. It was recognized very early \cite{oort46}, that in
cold environments the accretion of gas onto solid particles and their
agglomeration increases the number of large particles and decreases
the number of small ones. Grain destruction in hot environments has
the opposite effect, since large particles are shattered into smaller
ones and mass is removed from grains and returned to the gas phase by
sputtering. The effect of a variable mass distribution can be observed
by analyzing the wavelength dependence of the extinction of starlight
along different lines of sight through the interstellar medium (for a
review see Mathis, [1990]). Since the shape of the extinction curve is
sensitive to the mass distribution of dust grains that cause the
extinction, the mass distribution can be determined, in part, by
fitting the wavelength dependence of extinction
\cite{mathis77,kim94,li97}. These models of the mass distribution and
composition of interstellar dust are constrained by the total amount
of refractory elements available in the medium to form dust grains,
that is, the mass in refractory elements locked up in the dust plus the
mass in refractory elements in the gas phase should not exceed the
cosmic abundances of the elements. Consequently, models of
interstellar dust based on fits to the extinction curve do not contain
large amounts of big grains, because the large mass of refractory
elements contributed by the large dust grains would contradict cosmic
abundance considerations \cite{landgraf98a}. For example, the mass
distribution of the Mathis et al. [1977] model is cut off at grain
radii of $0.25\;{\rm \mu m}$ ($1.6\cdot 10^{-16}\;{\rm kg}$ assuming
a grain mass density of $\rho_{\rm d} = 2.5\cdot 10^3\;{\rm kg}\;{\rm
m}^{-3}$). In contradiction to this suggested cut-off, most of the grains
detected by Ulysses and Galileo have higher masses. Assuming that the
mass contained in the interstellar dust grains measured with Ulysses
and Galileo in the Solar System represents the value of the dust mass
density in the local interstellar cloud (LIC), and comparing it with
the mass contained in the gas phase of the LIC, it was found
\cite{frisch99} that the mass of refractory elements found in the LIC
exceeds cosmic abundance limits.

For each impact onto the dust detectors, the Ulysses and Galileo dust
instruments measure the velocity, mass, and direction of an impacting
grain \cite{gruen92}. Since the
detectors are mounted at angles of $95^\circ$ (Ulysses) and $60^\circ$
(Galileo) with respect to the spin-axes of the spacecraft, the impact
direction is given by the detector pointing (rotation angle) at the
time of impact with an accuracy of $\pm 70^\circ$ (field of view). The
signal rise time is a measure for the impact velocity and the mass of
the impacting grain is then derived from the impact velocity and the
signal amplitude. By calibrating the instruments on the ground
\cite{gruen95a} it was found that the derived mass depends strongly on
the impact velocity (see below). Earlier determinations of the mass
distribution of interstellar dust grains detected by Ulysses and
Galileo \cite{gruen94,landgraf96} used the masses derived from the
measured impact velocity, which can only be determined with an
uncertainty of a factor of $2$, resulting in an uncertainty of the
derived mass of a factor of $10$ \cite{gruen95a}. It was argued
\cite{landgraf98b} that the mass of the dust grains can be determined
with a higher accuracy if the impact velocity was determined from
theoretical considerations.

In this work, we determine the mass distribution of interstellar dust
grains detected by Ulysses and Galileo by using impact velocities
calculated from simple dynamical models. The resulting mass
distribution should be more reliable, provided that the 
\begin{figure*}
\epsfbox{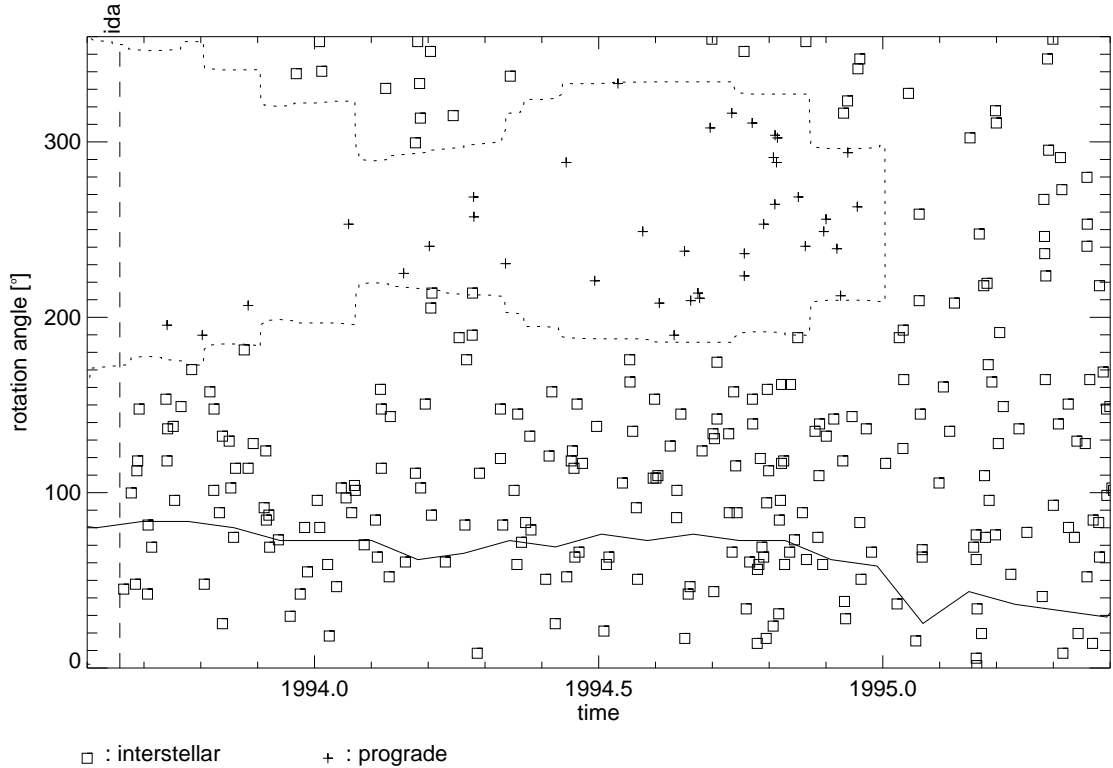}
\caption{\label{fig_galdata}\it Rotation-angle-vs-time plot of impacts
detected with the Galileo dust instrument. Crosses represent impacts
of particles that approached from the prograde direction. Impacts
that we identify as of interstellar origin (total number: $268$) are
shown as squares. The solid line gives the rotation angle at which the
maximum sensitive area is exposed to the interstellar upstream
direction (as defined by the stream direction of neutral Helium
\protect\cite{witte93}). The dotted contour contains the directions
and times when the upstream direction was in the field of view of the
instrument or the angle between the expected relative dust velocity
vector and the instrument pointing vector was less than
$80^\circ$. The time of the fly-by of the asteroid Ida is indicated by
the dashed vertical line.}
\end{figure*}

approximated
impact velocities deviate by less than a factor of $2$ from the true
values. To visualize the contribution of different grain sizes to the
number, cross section and total mass of dust grains in the LIC, we use
``moments'' of the differential mass distribution. The $i$-th moment
of the differential logarithmic distribution $n(m)\;d\log m$ is given
by $m^i\cdot n(m)\;d\log m$. In section \ref{sec_vimodel} we describe
the procedure of the mass determination and the models used to
calculate the impact velocity, section \ref{sec_massdis} describes the
resulting mass distributions and their moments. In section
\ref{sec_radar} we compare the Ulysses and Galileo measurements with
the measurements of recently discovered \cite{taylor96} interstellar
radar meteors.

\section{Dataset of Interstellar Dust Impacts on Ulysses and Galileo
Instruments \label{sec_datasets}}
For our analysis we use Galileo dust data collected between September 1993
(after fly-by of Asteroid Ida) and July 1995 (prior to probe release
operations). The Ulysses dataset we use contains impacts measured
between February 1992 (after Jupiter fly-by) and February 1999 (latest
data). Since the dust datasets collected by Ulysses and Galileo contain
impacts by interplanetary as well as interstellar grains, we have to
find selection criteria that allow us to define datasets, that contain
a negligible amount of other than interstellar impacts.

Both, Ulysses and Galileo have detected dust particles that had been
ejected from the Jovian system in streams \cite{gruen93}. These
streams can easily be identified in the datasets, because they occur
within a short period of time. Impacts of stream particles have been
removed from both datasets by rejecting impacts that occur in a time
interval that is associated with a stream event.

Interstellar particles approach Ulysses and Galileo from a direction
that is opposite to the direction expected for classic interplanetary
grains, that is grains on prograde, circular orbits. Therefore, we
preliminarily define interstellar datasets by selecting every impact
that was measured at a rotation angle of the sensor for which the
interstellar upstream direction was within the field of view of the
detector. We allow for a $10^\circ$ margin, because interstellar
grains do not move on perfectly straight lines. Since the sensor has a
field of view of $\pm 70^\circ$, we expect no particles on prograde
circular orbits to be present in the so defined datasets. We now
remove impacts from the preliminary datasets for which an other than
interstellar origin has been suggested. For Galileo, no interplanetary
source has been suggested for impacts from the retrograde direction
measured outside the asteroid belt. We therefore use the dataset of
$268$ impacts shown in figure~\ref{fig_galdata} for our
further analysis.
%Prior to the
%reprogramming of the Galileo instrument on 14 July 1994
%\cite{krueger98b}, events in the lowest coincidence class are
%suspected to be noise events. Therefore, we remove events from the
%Galileo dataset that have been classified to the lowest coincidence
%class prior to the reprogramming.

When Ulysses crossed the ecliptic at a heliocentric distance of
$1.3\;{\rm AU}$ in March 1995, the directions of interstellar and
prograde interplanetary grains were not as clearly separated as it was
the case during the rest of Ulysses' orbit. We therefore exclude all
measurements when Ulysses was between $-60^\circ$ and $+60^\circ$
ecliptic latitude around perihelion. Over the Sun's poles, Ulysses
detected very small particles that have been interpreted
\cite{hamilton96} as fragments of interplanetary particles ejected
from the inner Solar System by electromagnetic effects. To remove
these particles from the Ulysses dataset, we require that the measured
amplitude of the ion charge signal, which increases with impact
velocity and the mass of the impactor, is more than one order of
magnitude above the detection threshold. The described criteria are
identical to the criteria used by Landgraf [1998], but here they are
applied to a more recent dataset. After removing possible
interplanetary impacts from the dataset as described above, we use the
dataset of $444$ impacts shown in figure~\ref{fig_ulsdata} for our
analysis.

By changing the numerical values of the parameters we use to select
the datasets (directional margin, charge signal amplitude cut-off) we
estimate the relative uncertainty in the number of interstellar
particles to be smaller than $20\%$, on a $90\%$ confidence level.

\section{Theoretical Impact Velocities and Derived
Mass\label{sec_vimodel}} 

The change of velocity of an interstellar grain in the Solar System
can easily be determined by taking into account the acceleration due
to solar gravity and radiation pressure. For sub-micron grains the
strength of radiation pressure, expressed as the ratio of magnitudes
of radiation pressure to gravity $\beta$, can be in the same order
($\beta \approx 1$) or even larger ($\beta > 1$) as the strength of
gravity. In this work we use two simple models to determine the
velocity of the grains:
\begin{figure*}
\epsfbox{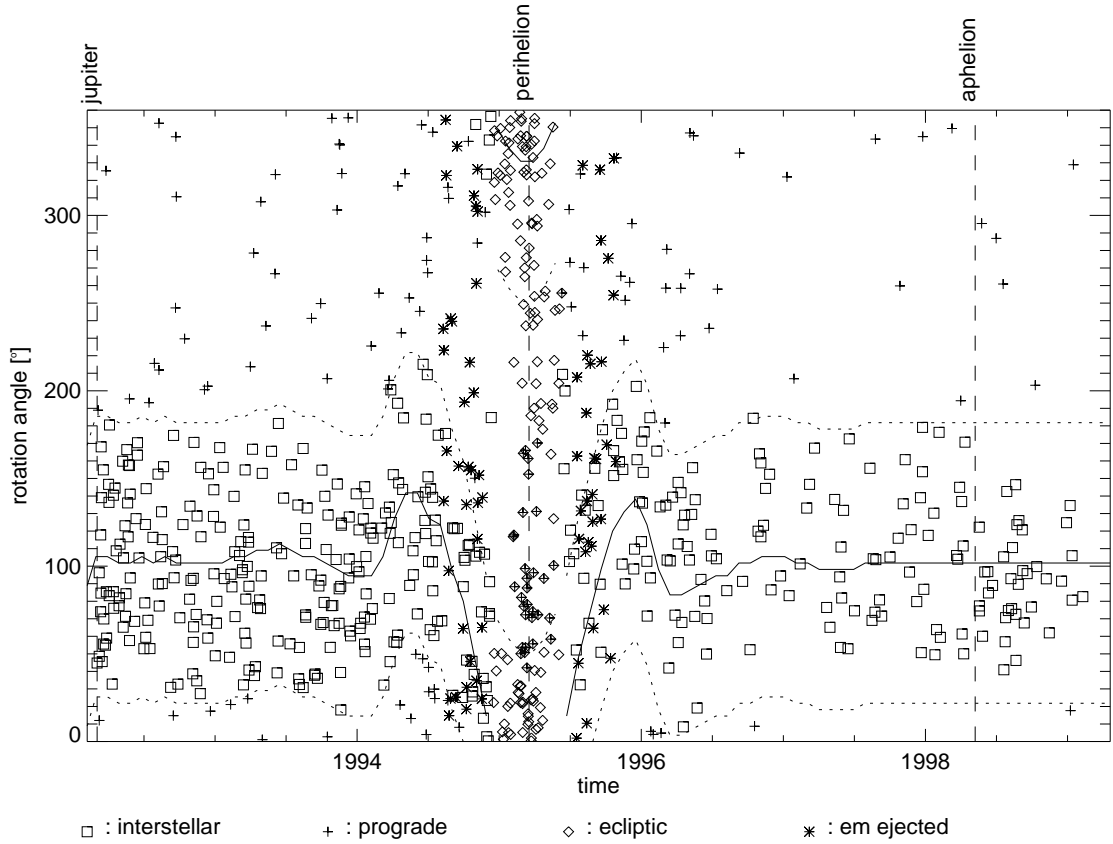}
\caption{\label{fig_ulsdata}\it Rotation-angle-vs-time plot of impacts
detected with the Ulysses dust instrument. Crosses represent impacts
of particles that approached from the prograde direction. As described
in the text, impacts around perihelion ecliptic crossing (diamonds)
and small amplitude impacts above the poles (stars) have been removed
from the dataset of interstellar impacts. Impacts that we identify as
of interstellar origin (total number: $444$) are shown as squares. The
solid line gives the rotation angle at which the maximum sensitive
area was exposed to the interstellar upstream direction (as defined
by the stream direction of neutral Helium \protect\cite{witte93}). The
dotted contour contains the directions and times when the upstream
direction was in the field of view of the instrument or the angle
between the expected relative dust velocity vector and the instrument
pointing vector was less than $80^\circ$. The times of the Jupiter
fly-by, and the perihelion and aphelion ecliptic crossings are
indicated by dashed vertical lines.}
\end{figure*}

\begin{figure}
\epsfxsize=.9\hsize
\epsfbox{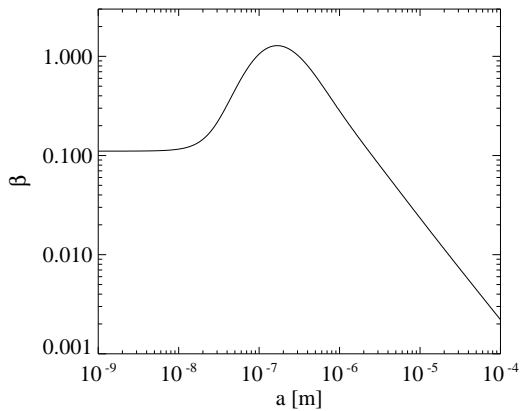}
\caption{
\figurewidth{.45\hsize}
\label{fig_bobeta}\it The ratio  $\beta$ of magnitudes of radiation
pressure force to gravity as a function of grain radius $a$ as given
by Gustafson [1994]. The optical properties of Astronomical Silicates,
a spherical shape and a bulk density of $2.5\cdot 10^3\;{\rm kg}\;{\rm
m}^{-3}$ have been assumed.}
\end{figure}

\begin{enumerate}
\item The radiation pressure force and gravity have exactly the same
magnitude and opposite directions ($\beta = 1$). Therefore, the grains
move on straight lines with their initial velocity and direction. In
this case the impact velocity is simply given by the difference of the
grain velocity at infinity and the spacecraft velocity.
\item The ratio $\beta$ of radiation pressure force to gravity is
given by the grain size. The velocity $v_{\rm ECL}$ of the dust grain
in the inertial (heliocentric, ecliptic) frame is changed by the
acceleration ($\beta < 1$) or deceleration ($\beta > 1$) along its
trajectory according to
\begin{eqnarray}
v_{\rm ECL} & = & \sqrt{v_\infty^2 + \frac{2\gamma (1 - \beta)
M_\odot}{r_{\rm hc}}},\label{eqn_vecl}
\end{eqnarray}
where $\gamma$ is the gravitational coupling constant, $M_\odot$ the
mass of the Sun, and $r_{\rm hc}$ the heliocentric distance at which the
grain was detected.
\end{enumerate}
For the calculation of the heliocentric grain velocity as described in
the second model, we have to determine the value of $\beta$ for each
individual grain. To determine $\beta$ we use Mie-calculations
\cite{gustafson94} for compact spherical grains made of astronomical
silicates with bulk densities of $2.5\cdot 10^3\;{\rm kg}\;{\rm
m}^{-3}$, which give $\beta$ as a function of grain radius (see
figure~\ref{fig_bobeta}). 
\vspace{1mm}
Because
the size $a$ of a grain is not measured independently by the
Ulysses and Galileo dust detectors, it has to be derived from the measured
mass $m_{\rm meas}$, i.e. the mass derived from the measured impact
velocity, by $a=\sqrt[3]{3m_{\rm meas}/(4\pi\rho_{\rm d})}$.
Using the grain radius $a$, we can determine $\beta$ which gives us the
dust velocity $v_{\rm ECL}$ from equation (\ref{eqn_vecl}).
Once the the dust velocity in the inertial heliocentric frame is
established, the impact velocity $v_i$ is calculated as the magnitude
of the velocity relative to the spacecraft. The mass is then given by
\begin{eqnarray}
m[{\rm g}] & = & 1.7\cdot 10^{-5}\cdot Q_I[{\rm C}]\cdot
\left(v_i[{\rm km}\;{\rm s}^{-1}] \right)^{-3.5},\label{eqn_calib}
\end{eqnarray}
as described by Gr\"un et al. [1995], where $Q_I$ is the amplitude of
the measured ion charge signal. From this mass we determine a new
grain size which gives a new $\beta$, and so forth. This iterative
process gives us the value of $\beta$ and the mass of a particle in a
self-consistent way. A disadvantage of this second model is its
dependence on not well known properties of presumably complex dust
grains. Therefore, we rely on the first (constant velocity) model for
grains for which $\beta=1$ is a good approximation. It was found that
the constant-velocity model is a good approximation for the majority
of the impacts detected by Ulysses and Galileo, because the
Mie-calculations give a mean value of $\beta$ for all particles
measured in-situ is $\overline{\beta}=1.02$, and for $90\%$ of the
detections $\beta$ deviates by less than $0.6$ from unity
\cite{landgraf98b}. But since we are interested in the large mass end of
the distribution, where $\beta=1$ is not a good approximation, we
apply both methods to the datasets collected by Ulysses and Galileo,
and compare the results.

For both models described above, we assume an initial velocity of
$26\;{\rm km}\;{\rm s}^{-1}$ and an upstream direction of $259^\circ$
heliocentric longitude and $8^\circ$ heliocentric latitude. This
initial velocity vector is: (a) compatible with the heliocentric
speeds \cite{gruen94} and direction of motion \cite{baguhl95a} of the
grains detected by Ulysses, (b) close to the asymptotic velocity
vector of interstellar neutral Helium atoms detected by the
Ulysses/GAS experiment \cite{witte93}, and (c) compatible with the
relative velocity of the Sun with respect to the LIC
\cite{lallement92}. We neglect the Lorentz-force exerted on the grains
by interaction with the solar wind magnetic field
\cite{landgraf99}. This is a good approximation as long as the
direction of motion is not strongly changed by the Lorentz-force,
which is true for particles with masses larger than $10^{-17}\;{\rm
kg}$, since their Larmor-radii are in the order of $500\;{\rm AU}$
\cite{gruen94}, much larger than the length
of their path through the domain of the solar wind.

\section{Mass Distributions \label{sec_massdis}}
We determine the grain mass distributions and their moments of the
datasets described in section~\ref{sec_datasets}. 

\begin{figure*}
\begin{tabular}{cc}
\epsfxsize=.5\hsize
\epsfbox{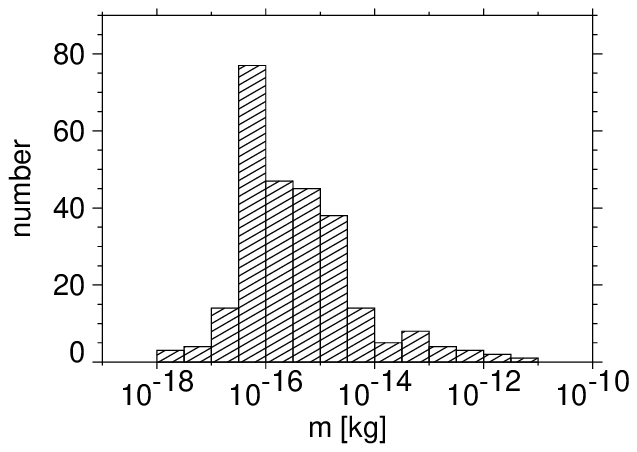} &
\epsfxsize=.5\hsize
\epsfbox{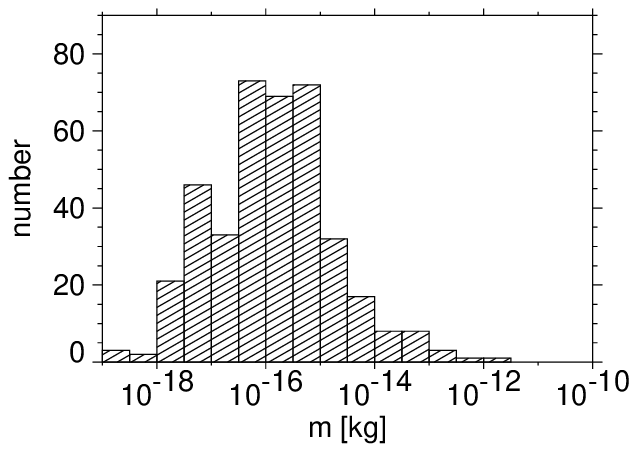}
\end{tabular}
\caption{\label{fig_mh_meas}
\it Histograms of the mass of the
interstellar grains detected by Galileo (left panel) and Ulysses
(right panel). The masses have been derived from the measured impact
velocities.}
\end{figure*}

\begin{figure*}
\begin{tabular}{cc}
\epsfxsize=.5\hsize
\epsfbox{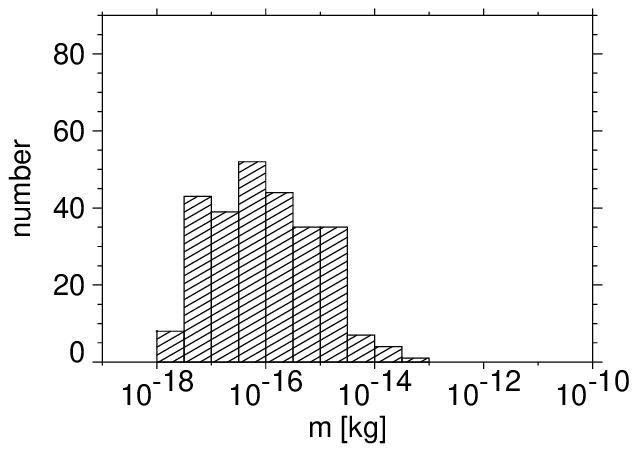} &
\epsfxsize=.5\hsize
\epsfbox{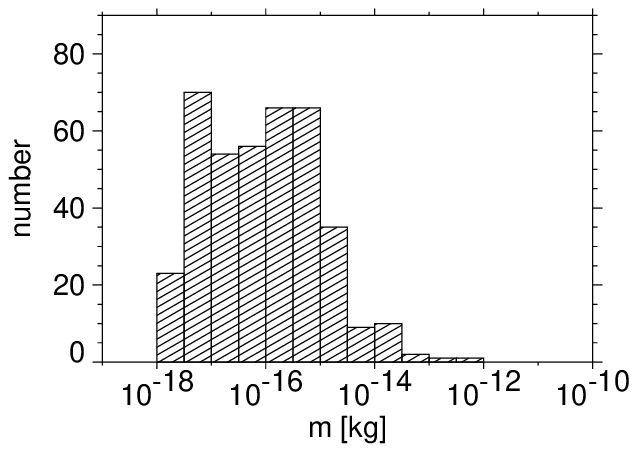}
\end{tabular}
\caption{\label{fig_mh_beq1} 
\it Histograms of the mass of the
interstellar grains detected by Galileo (left panel) and Ulysses
(right panel). The masses have been derived from the straight-line
($\beta=1$) model.}
\end{figure*}

\begin{figure*}
\begin{tabular}{cc}
\epsfxsize=.5\hsize
\epsfbox{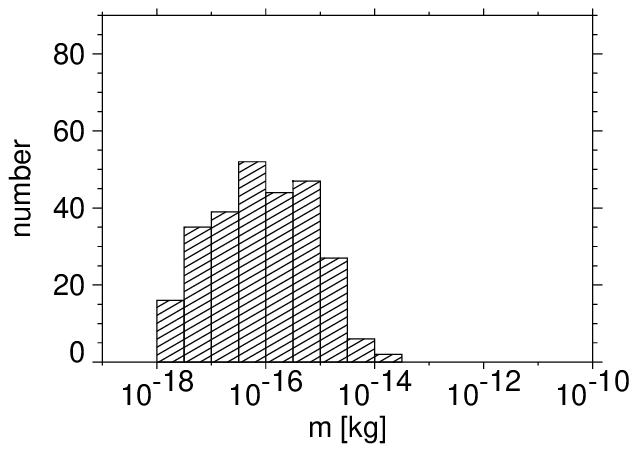} &
\epsfxsize=.5\hsize
\epsfbox{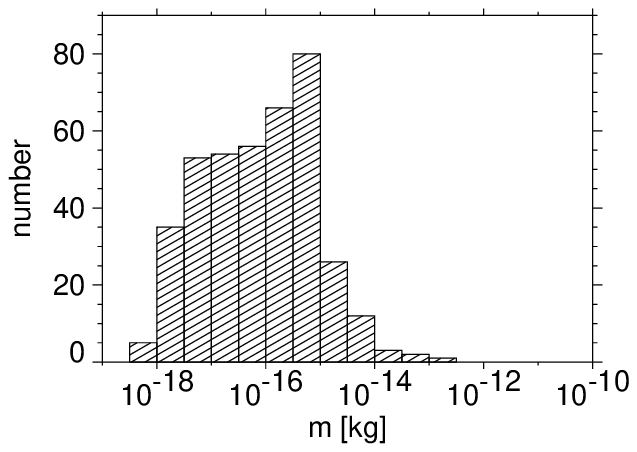}
\end{tabular}
\caption{\label{fig_mh_bne1}\it Histograms of the mass of the
interstellar grains detected by Galileo (left panel) and Ulysses
(right panel). The masses have been derived from the self consistent
model of accelerated (decelerated) grains with $\beta<1$ ($\beta >
1$).}
\end{figure*}

The histograms of the distributions of grain masses that have been
derived from the measured impact velocities are shown in
figure~\ref{fig_mh_meas} (compare Gr\"un et al. [1993], Gr\"un et
al. [1995], Baguhl et al. [1996], Landgraf and Gr\"un [1998]).  The
distributions cover a mass interval from $10^{-18}\;{\rm kg}$ (which
is the detection threshold of grains impacting with $20\;{\rm
km}\;{\rm s}^{-1}$) to $10^{-12}\;{\rm kg}$ and peak at
$10^{-16}\;{\rm kg}$. From modeling the extinction of starlight
\cite{mathis77}, the number of grains per mass interval is expected to
increase steeply for smaller masses. In contradiction to this
expectation, the in-situ measurements indicate a deficiency of grains
in the lower mass region between $10^{-18}\;{\rm kg}$ and
$10^{-16}\;{\rm kg}$. This deficiency has been interpreted as being
due to the electro-magnetic interaction of the grains with the solar
wind magnetic field \cite{gruen94,landgraf99}. The upper limit of the
grain mass range is determined by the limited size of the Ulysses and
Galileo. Since large grains are much less abundant than small ones, no
statistically significant number of particles with masses larger than
$10^{-12}\;{\rm kg}$ was detected.  The distribution of masses
determined by assuming a constant dust velocity of $26\;{\rm km}\;{\rm
s}^{-1}$ and straight trajectories (first model described in section
\ref{sec_vimodel}, $\beta=1$) are shown in
figure~\ref{fig_mh_beq1}. Comparing these histograms with
the distributions shown in figure~\ref{fig_mh_meas}, we find that the
number of particles with small masses ($m < 10^{-16}\;{\rm kg}$) is
increased and the number of particles with large masses ($m >
10^{-16}\;{\rm kg}$) is decreased when we derive the grain masses from
the constant-velocity model instead from measured impact
velocities. This is because the measured impact velocities of large
grains are systematically lower than assumed impact velocity of
interstellar particles at the location of the measurement. This can
have two possible explanations: (a) The datasets contain unidentified
interplanetary (bound) grains that impact the detector with a velocity
lower than the assumed hyperbolic velocity, and (b) the measurements
of velocities of fast and big grains deviate systematically from the
true value to lower values. Such a systematic deviation can occur if
recombinations take place in the plasma cloud that is generated by the
impact, if the plasma density is high enough. In
figure~\ref{fig_mh_bne1} we show the resulting histogram of masses
derived from impact velocities that have been calculated taking into
account the acceleration ($\beta<1$) or deceleration ($\beta > 1$) of
grains by the combined forces of solar gravity and radiation
pressure. The resulting distribution is similar to the result obtained
by neglecting the acceleration, but the number of grains with large
masses is further reduced, because we used more realisticly higher
impact velocities (see equation (\ref{eqn_calib})) were used to derive
the grain mass.

As mentioned in the introduction and discussed in Frisch et
al. [1999], the existence of interstellar grains with masses much
larger than $10^{-16}\;{\rm kg}$ is an important result of the in-situ
dust measurements of Ulysses and Galileo. These large grains are not
expected to contribute significantly to the total optical cross section
of the interstellar dust population and can thus be difficult to
observe. In the following we 

\begin{figure*}
\epsfxsize=\hsize
\epsfbox{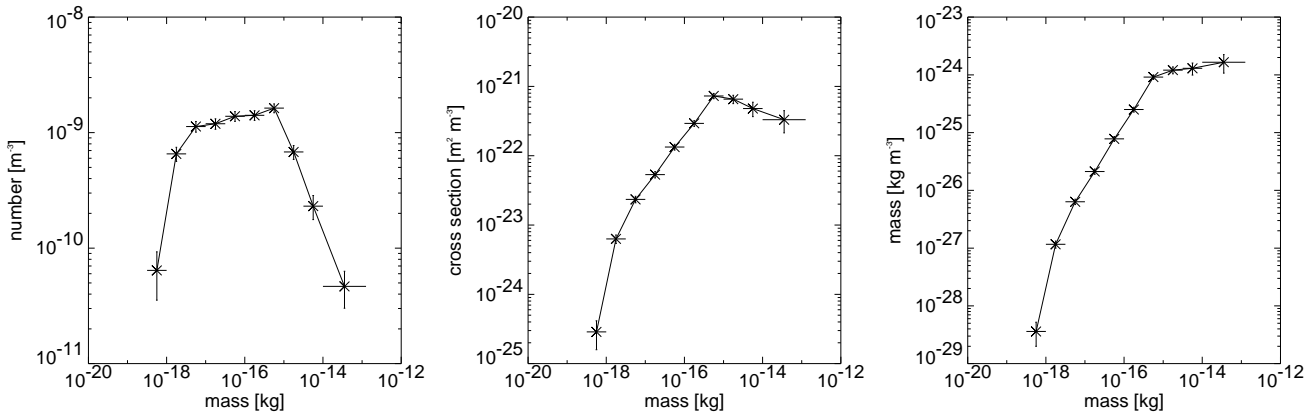}
\caption{\label{fig_moments}\it Moments of the differential mass
distribution of interstellar grains measured by Ulysses and Galileo in
the Solar System. The panels show the number (left), cross section
(middle), and mass (right) per logarithmic mass interval and unit
volume.}
\end{figure*}

calculate the contribution of grains of
different masses to the concentration, cross section per unit volume,
and total mass density of the interstellar dust population. Here we
use the masses that we have derived from calculated impact velocities,
taking into account acceleration by solar gravity and radiation
pressure. To gain better statistics, we combine the Ulysses and
Galileo measurements. Since both detectors had different exposure
times to the interstellar upstream direction, we calculate number of
impacts per logarithmic mass interval and unit volume for both
datasets, and combine them by calculating the geometric average. The
resulting differential number density mass distribution $n(m)\;d\log
m$ is shown in the left panel of figure~\ref{fig_moments}.
The number density is dominated by grains with masses between
$10^{-17}\;{\rm kg}$ and $10^{-15}\;{\rm kg}$, as expected from the
histograms of the Ulysses and Galileo data.

The cross-section-mass distribution is given by
\begin{eqnarray}
\sigma(m)\cdot n(m)\;d\log m & = & \pi a^2(m) n(m)\;d\log m \nonumber\\
& = & \pi \left(
\frac{3 m}{4\pi \rho_{\rm d}} \right)^\frac{2}{3} n(m)\;d\log m.\nonumber\\
\end{eqnarray}
The resulting distribution of the combined Ulysses and Galileo data is
shown in the middle panel of figure \ref{fig_moments}. In the in-situ
data, the grain masses between $10^{-21}\;{\rm kg}$ and
$10^{-16}\;{\rm kg}$, that are believed to cause the extinction of
starlight, do not contribute dominantly to the cross section. The
reason is that their number is depleted in the Solar System by their
interaction with the solar wind (see description of the histograms,
figure \ref{fig_mh_meas}). Therefore, Ulysses and Galileo did not
detect abundantly the grains that cause the extinction of starlight.

The contribution of grains of different masses to the overall mass
density of interstellar dust can be represented by the distribution of
mass density per logarithmic mass interval $m\cdot n(m)\; d\log m$
(see right panel of figure \ref{fig_moments}). For the total mass
density, the biggest particles measured become important. Calculating
the total mass density from the Ulysses and Galileo in-situ data by
integrating over the differential mass-density-mass distribution, we
get the result $6.2\cdot 10^{-24}\;{\rm kg}\;{\rm m}^{-3}$.

From the value for the total mass density of in-situ detected dust
grains we can extrapolate the gas-to-dust mass ratio in the LIC, which
gives us information about the amount of refractory elements in the
local interstellar environment. Adopting a Hydrogen density of $n_{\rm
H} = 3\cdot 10^{5}\;{\rm m}^{-3}$ and a Helium density of $n_{\rm He}
= 3\cdot 10^{4}\;{\rm m}^{-3}$ given by Frisch et al. [1999], and
using the total mass density of interstellar grains detected by
Ulysses and Galileo, the gas-to-dust mass ratio in the LIC is
$113$. This figure lies within the $1\sigma$-range of the value reported by
Frisch et al. [1999] of $94^{+46}_{-38}$. As can be seen from the
right panel in figure \ref{fig_moments}, the mass contribution of a
given mass interval increases monotonically with mass. Therefore, the
given total mass density is a lower estimate of the true value,
because the upper mass limit of the integration is defined by the
largest impact detected. Since the total number of large particles is
low (left panel of figure \ref{fig_moments}), the upper limit of the
mass integration depends strongly on the sensitive area of the
instrument and the accumulated time of measurements. To estimate the
total mass density of grains in the LIC more realisticly, we need
information about the number of grains larger than the largest grain
detected by Ulysses and Galileo. This data can be provided by the
measurement of hyperbolic radar meteors described in section
\ref{sec_radar}.

\begin{figure}
\epsfxsize=.9\hsize
\epsfbox{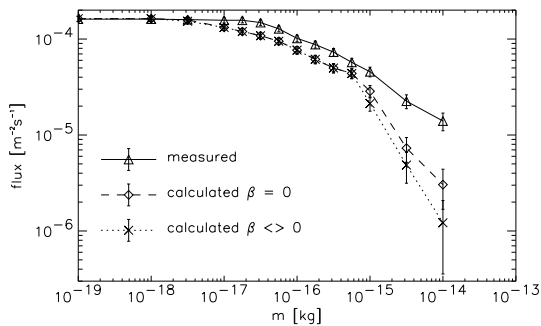}
\caption{\label{fig_cumflux_g} 
\figurewidth{.45\hsize}
\it The cumulative flux mass distribution
of interstellar grains detected by Galileo using three methods to
determine the impact velocity for the mass calibration: measured by
the instrument (triangles connected by solid line), particles with
$\beta=1$ move on straight lines with $v_\infty = 26\;{\rm km}\;{\rm
s}^{-1}$ (diamonds connected by dashed line), and particles with
$\beta\neq 1$ that are accelerated by solar gravity and radiation
pressure with a $\beta$-value according to their size (crosses connected
by dotted line).}
\end{figure}

\begin{figure}
\epsfxsize=\hsize
\epsfbox{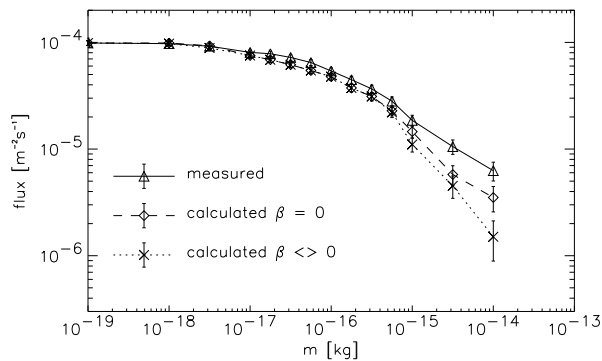}
\caption{\label{fig_cumflux_u} 
\figurewidth{.45\hsize}
\it The same distributions as shown in figure
\ref{fig_cumflux_g} for the Ulysses dataset.}
\end{figure}

As described above, the mass we derive for big interstellar grains
depends significantly on the impact velocity we assume. This effect
can be seen in figures~\ref{fig_cumflux_g} and~\ref{fig_cumflux_u}
that show the cumulative-flux-mass distribution of interstellar grains
detected by Galileo and Ulysses, respectively. 


For both datasets, the
flux of large grains from the interstellar direction is overestimated
when using the measured impact velocity to derive a mass. As mentioned
above, this can be due to either misidentification of interplanetary
grains or to systematic deviation of the measured impact velocity to
lower values. Since the flux of large grains from the interstellar
direction measured by Galileo is higher than the value derived from
the Ulysses data, and Galileo operated in the ecliptic plane, where
interplanetary grains on highly eccentric orbits could enter the
detector from the retrograde direction, we conclude that the Galileo
dataset contains a component of large interplanetary grains. Since a
predominantly retrograde population of interplanetary grains is not
believed to exist in the Solar System, and large impacts detected by
Ulysses outside the ecliptic plane came from the retrograde direction
(which coincides with the interstellar upstream direction), we
conclude that the Ulysses dataset does not contain a significant
contamination by an interplanetary component, and that the flux of
interstellar grains of masses above $10^{-14}\;{\rm kg}$ is in the
order of $10^{-6}\;{\rm m}^{-2}\;{\rm s}^{-1}$.

\begin{figure*}
\begin{tabular}{cc}
\epsfxsize=.5\hsize
\epsfbox{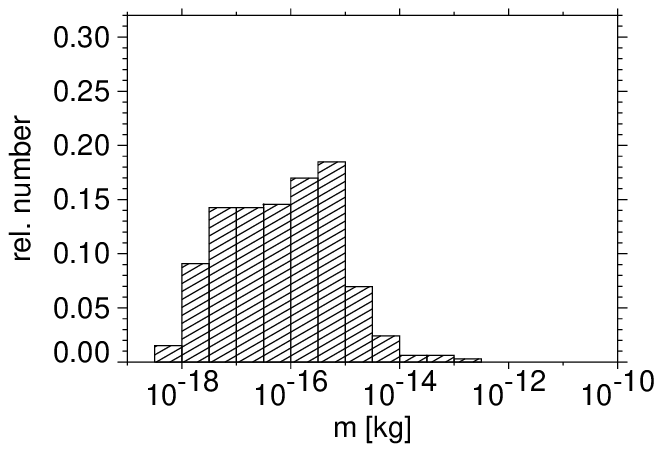} &
\epsfxsize=.5\hsize
\epsfbox{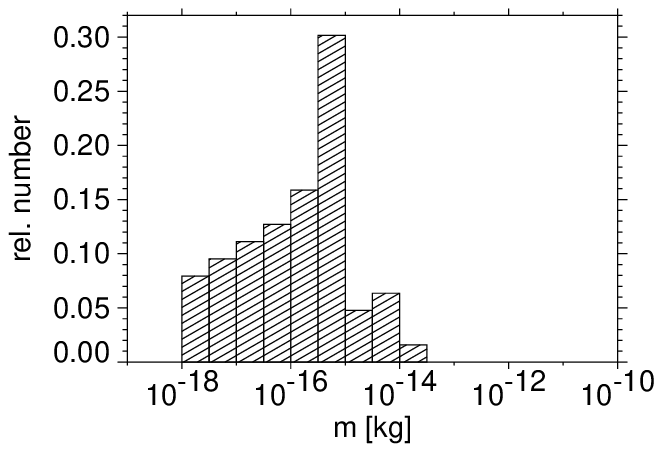}
\end{tabular}
\caption{\label{fig_mh_earlyvslate} 
\it
Histograms (relative abundance) of
the mass of interstellar grains detected by Ulysses before (left) and
after (right) April 1996. The masses have been derived from the self
consistent model of accelerated (decelerated) grains with $\beta<1$
($\beta > 1$). The later mass distribution contains less small
($m < 10^{-16}\;{\rm kg}$) grains.}
\end{figure*}

Modeling of the interaction of small interstellar grains with the
solar wind magnetic field \cite{landgraf99} suggests that the mass
distribution changes with time. One result of the modeling is that
small grains (with radii of $\approx 0.2\;{\rm \mu m}$) are depleted
after mid 1996 because of the diverting configuration of the solar
wind magnetic field. The analysis of the mass distribution of grains
detected by Ulysses before and after April 1996 indicates such a
depletion as shown in figure~\ref{fig_mh_earlyvslate}. The
ratio of the number of particles with masses lower than
$10^{-16}\;{\rm kg}$ to the number of particles with masses larger
than $10^{-16}\;{\rm kg}$ was $1.2\pm 0.18$ before, and $0.7 \pm 0.25$
after April 1996 when using calculated impact velocities to determine
grain masses (when using measured impact velocities, the corresponding
numbers are: $0.88\pm 0.14$ before, and $0.66\pm 0.24$ after April
1996). Unfortunately, the change in the ratio is not statistically
significant and more data is needed to prove or disprove the suggested
time dependence of the mass distribution.

\section{Large Grain Masses and the Flux of Large Interstellar Dust
 Grains Detected as Radar Meteors \label{sec_radar}} 

The existence of large ($m > 10^{-15}\;{\rm kg}$) interstellar grains
is a possible explanation for the observation that the grain
population in the interstellar 

medium is replenished faster than by
condensation in stellar outflows only \cite{jones94}. It is argued
\cite{gruen99} that large grains have long lifetimes in the
interstellar medium and provide a source for smaller grains which can
then be observed in extinction. Since the Ulysses and Galileo
measurements provide limited statistics of the number of large grains,
we compare detections of interstellar radar meteors with Ulysses and
Galileo data to determine how and if the mass distribution of
interstellar dust grains detected by the spacecraft can be
extrapolated to higher masses.

Since the measurements by AMOR \cite{baggaley99} allow the
determination of the impact trajectory and velocity of meteoroids in
the Earth's atmosphere, hyperbolic meteoroids can be identified. It is
then possible to determine the flux of large ($m > 3\cdot
10^{-10}\;{\rm kg}$) interstellar meteors for various
source directions. It was found that the flux exhibits an ecliptic
north-south asymmetry. The flux from the northern ecliptic
hemisphere (including the upstream direction of grains detected by
Ulysses and Galileo) was reported to be lower than $3\cdot
10^{-10}\;{\rm m}^{-2}\;{\rm s}^{-1}$, whereas a flux of $1.8\cdot
10^{-8}\;{\rm m}^{-2}\;{\rm s}^{-1}$ was reported from the southern
ecliptic hemisphere. A discrete source direction of large interstellar
grains could be identified for which a flux of $2\cdot 10^{-9}\;{\rm
m}^{-2}\;{\rm s}^{-1}$ was determined \cite{baggaley99}.

\begin{figure*}
\centering
\epsfxsize=.6\hsize
\epsfbox{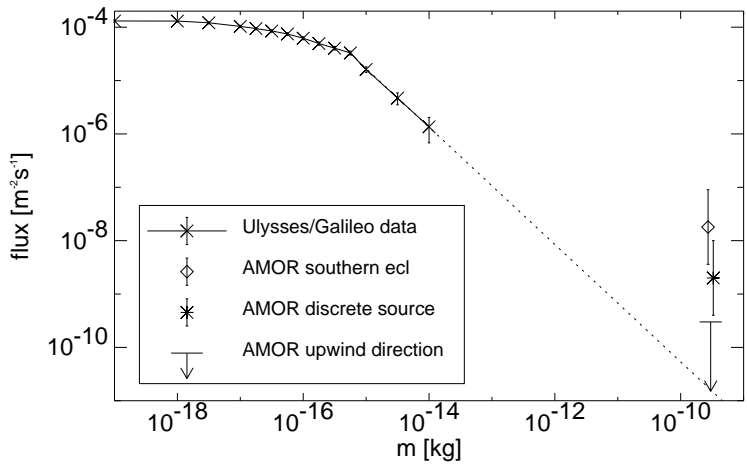}
\caption{\label{fig_radarcomp} 
\it The cumulative flux mass distribution
of the Ulysses and Galileo data compared to the flux of interstellar
radar meteors detected with AMOR. The dotted line represents an
extrapolation by a power-law with an exponent of $-1.1$. The radar
data for different source directions is shown as a diamond (influx from
southern ecliptic hemisphere), a star (influx from the discrete
source), and as a horizontal bar indicating an upper limit (influx
from the upstream direction of grains detected by Ulysses and
Galileo).}
\end{figure*}

The comparison of the large-grain-flux derived from the radar
measurements with the extrapolation of the Ulysses and Galileo results
is shown in figure~\ref{fig_radarcomp}.  We extrapolate the
cumulative Ulysses and Galileo mass distribution by fitting a
power-law function to the distribution of masses larger than $5\cdot
10^{-16}\;{\rm kg}$, where the distribution is not affected by
electromagnetic effects. The exponent of the resulting fit function is
$-1.1 \pm 0.1$. From the extrapolation of the Ulysses and Galileo mass
distribution, we expect a flux of less than $10^{-10}\;{\rm
m}^{-2}\;{\rm s}^{-1}$ for grains with masses larger than $3\cdot
10^{-10}\;{\rm kg}$. This is compatible with the upper limit given for
interstellar radar meteors coming from the same direction as the
Ulysses and Galileo particles. The flux of interstellar radar meteors
from the southern ecliptic hemisphere as well as from the discrete
source is one or two orders of magnitude larger than the value
expected from the extrapolation of the Ulysses and Galileo
measurements.

\section{Conclusion}
We have derived a mass distribution of interstellar grains from the
in-situ data gathered by the Ulysses and Galileo dust detectors using
the assumption that interstellar grains impact the detectors with
velocities that are given by their initial velocity of $26\;{\rm
km}\;{\rm s}^{-1}$ at large heliocentric distances, the acceleration
by solar gravity and radiation pressure, and the motion of the
spacecraft. As a result we find that the number of large particles
decreases and the number of intermediate size particles increases when
deriving masses from calculated instead of measured impact
velocities. However, we find that the values derived for 

total mass density
and the gas-to-dust mass ratio in the LIC do not change significantly
from the values reported by Frisch et al. [1999].

In the Ulysses data we find an indication that the mass distribution
of interstellar grains is time-variable, as predicted by model
calculations \cite{landgraf99}. Long-term measurements are needed
to assess the significance of this variability.

Gr\"un and Landgraf [1999] suggest that the interstellar mass distribution
in the LIC consists of two parts: Grain masses below $10^{-16}\;{\rm
kg}$ are distributed approximately like the dust component that causes
extinction in the diffuse interstellar medium \cite{mathis77}. At
$10^{-16}\;{\rm kg}$ a transition to a steeper mass distribution is
suggested such that the large grain component still contains a
significant amount of mass, but the contribution of each mass decade
does not increase further for increasing grain mass. We determine the
slope of the logarithmic cumulative in-situ flux-mass distribution to
be $-1.1$. This figure is close to the slope of $-1.0$, for which each
mass decade contributes the same amount to the total mass.
Unfortunately, the shape of the measured mass distribution below
$10^{-16}\;{\rm kg}$ is strongly affected by heliospheric filtration
\cite{landgraf99} and can therefore not be compared directly to the
expected mass distribution. In summary we find that the in-situ
measurements support the interpretation of an extended mass
distribution as described by Gr\"un and Landgraf [1999].

Radar data collected by the AMOR facility in New Zealand indicates
that there is a large reservoir of mass in big interstellar grains in
the solar vicinity. Because these large grains couple to the gas phase
of the interstellar medium on length scales much larger than the size
of the LIC \cite{morfill79a}, these big grains are not believed to be
related to the grain population detected by Ulysses and Galileo, which
enters the Solar System from the same direction as Helium atoms that
originate from the gas phase of the LIC. The radar measurements
indicate that the majority of radar meteors arrives from the southern
ecliptic hemisphere. If they were accompanied by a large number of
smaller grains, Ulysses and Galileo would have detected them, which is
not the case. Since collisional evolution of large dust grains into
smaller ones is believed to be an effective process in the diffuse
interstellar medium\cite{jones96}, one can speculate that the grains
detected by AMOR have been accompanied by smaller grains, but the
small grains have been stopped on its way to the Solar System by an
interstellar cloud.

\acknowledgements
This work was performed while M.L. held a National Research
Council-NASA/JSC Research Associateship.

\end{article}

\begin{thebibliography}{}

\bibitem[{\it Baggaley,} 1999]{baggaley99}
\reference
W.~J.~Baggaley.
\newblock AMOR Radar Measurements of Interstellar Meteoroids.
\newblock {\em Journal of Geophysical Research}, this issue, 1999.

\bibitem[{\it Baguhl et al.,} 1995]{baguhl95a}
\reference
M.~Baguhl, D.~P. Hamilton, E.~Gr{\"u}n, S.~F. Dermott, H.~Fechting, M.~S.
  Hanner, J.~Kissel, B.-A. Lindblad, D.~Linkert, G.~Linkert, I.~Mann, J.~A.~M.
  McDonnell, G.~E. Morfill, C.~Polanskey, R.~Riemann, G.~Schwehm, P.~Staubach,
  and H.~A. Zook.
\newblock Dust measurements at high ecliptic latitudes.
\newblock {\em Science}, 268:1016--1019, 1995.

\bibitem[{\it Baguhl et al.,} 1996]{landgraf96}
\reference
M.~Baguhl, E.~Gr{\"u}n, and M.~Landgraf.
\newblock In situ measurements of interstellar dust with the Ulysses and
  Galileo spaceprobes.
\newblock In R.~von Steiger, R.~Lallement, and M.~A. Lee, editors, {\em The
  Heliosphere in the Local Interstellar Medium}, volume~78 of {\em Space
  Science Reviews}, pages 165--172. Kluwer Academic Publishers, 1996.

\bibitem[{\it Frisch et al.,} 1999]{frisch99}
\reference
P.~C. Frisch, J.~Dorschner, J.~Gei{\ss}, J.~M. Greenberg, E.~Gr{\"u}n,
  M.~Landgraf, P.~Hoppe, A.~P. Jones, W.~Kr{\"a}tschmer, T.~J. Linde, G.~E.
  Morfill, W.~Reach, J.~Slavin, J.~Svestka, A.~Witt, and G.~P. Zank.
\newblock Dust in the local interstellar wind.
\newblock {\em ApJ},
\newblock in press, 1999.

\bibitem[{\it Gr\"un et al.,} 1992]{gruen92}
\reference
E.~Gr\"un, H.~Fechtig, M.S.~Hanner, J.~Kissel, B.-A.~Lindblad,
D.~Linkert, D.~Maas, G.E.~Morfill, and H.A.~Zook.
\newblock The Galileo Dust Detector.
\newblock {\em Space Sci. Rev.}, 60:317--340, 1992.

\bibitem[{\it Gr\"un et al.,} 1993]{gruen93}
\reference
E.~Gr{\"u}n, H.~A. Zook, M.~Baguhl, A.~Balogh, S.~J. Bame, H.~Fechtig,
  R.~Forsyth, M.~S. Hanner, M.~Horanyi, J.~Kissel, B.-A. Lindblad, D.~Linkert,
  G.~Linkert, I.~Mann, J.~A.~M. McDonnell, G.~E. Morfill, J.~L. Phillips,
  C.~Polanskey, G.~Schwehm, N.~Siddique, P.~Staubach, J.~Svestka, and
  A.~Taylor.
\newblock Discovery of jovian dust streams and interstellar grains by the
  {U}lysses spacecraft.
\newblock {\em Nature}, 362:428--430, 1993.

\bibitem[{\it Gr\"un et al.,} 1994]{gruen94}
\reference
E.~Gr{\"u}n, B.~{\AA}.~S. Gustafson, I.~Mann, M.~Baguhl, G.~E. Morfill,
  P.~Staubach, A.~Taylor, and H.~A. Zook.
\newblock Interstellar dust in the heliosphere.
\newblock {\em Astronomy and Astrophysics}, 286:915--924, 1994.

\bibitem[{\it Gr\"un et al.,} 1995]{gruen95a}
\reference
E.~Gr{\"u}n, M.~Baguhl, H.~Fechtig, J.~Kissel, D.~Linkert, G.~Linkert, and
  R.~Riemann.
\newblock Reduction of galileo and ulysses dust data.
\newblock {\em Planet. Space Sci.}, 43:941--951, 1995.

\bibitem[{\it Gr\"un and Landgraf,} 1999]{gruen99}
\reference
E.~Gr{\"u}n and M.~Landgraf.
\newblock Collisional Consequences of Big Interstellar Grains.
\newblock {\em Journal of Geophysical Research}, this issue, 1999.

\bibitem[{\it Gustafson,} 1994]{gustafson94}
\reference
B.~{\AA}.~S. Gustafson.
\newblock Physics of zodiacal dust.
\newblock {\em Annual Review of Earth and Planetary Science}, 22:553--595,
  1994.

\bibitem[{\it Hamilton et al.,} 1996]{hamilton96}
\reference
D.~P. Hamilton, E.~Gr{\"u}n, and M.~Baguhl.
\newblock Electromagnetic escape of dust from the solar system.
\newblock In B.{\AA}.~S. Gustafson and M.~S. Hanner, editors, {\em Physics,
  Chemistry, and Dynamics of Interplanetary Dust}, volume 104 of {\em
  Astronomical Society of the Pacific Conference Series}, pages 31--34, 1996.

\bibitem[{\it Jones et al.,} 1994]{jones94}
\reference
A.~P. {Jones}, A.~G. G.~M. {Tielens}, D.~J. {Hollenbach}, and C.~F. {McKee}.
\newblock Grain destruction in shocks in the interstellar medium.
\newblock {\em Astrophysical Journal}, 433:797--810, 1994.

\bibitem[{\it Jones et al.,} 1996]{jones96}
\reference
A. P. Jones, A. G. G. M. Tielens, and D. J. Hollenbach.
\newblock Grain Shattering in Shocks: The Interstellar Grain Size
Distribution.
\newblock {\em Astrophysical Journal}, 469:740, 1996.

\bibitem[{\it Kim et al.,} 1994]{kim94}
\reference
S.-H. Kim., P.~G. Martin, and P.~D. Hendry.
\newblock The size distribution of interstellar dust particles as determined
  from extinction.
\newblock {\em Astrophysical Journal}, 422:164--175, 1994.

\bibitem[{\it Kr\"uger et al.,} 1998a]{krueger98a}
\reference 
H. Kr{\"u}ger, E. Gr{\"u}n, M. Landgraf, M. Baguhl, S. Dermott,
H. Fechtig, B.{\AA}. S. Gustafson, D. P. Hamilton, M. S. Hanner,
M. Hor\'anyi, J. Kissel, B. A. Lindblad, D. Linkert, G. Linkert,
I. Mann, J. A. M. McDonnell, G. E. Morfill, C. Polanskey, G. Schwehm, 
R. Srama, and H. A. Zook.
\newblock Three years of Ulysses dust data: 1993 to 1995.
\newblock to appear in {\em Planetary and Space Science}.

\bibitem[{\it Kr\"uger et al.,} 1998b]{krueger98b}
\reference
H. Kr{\"u}ger, D. P. Hamilton, M. Baguhl, S. Dermott, H. Fechtig, 
B.{\AA}. S. Gustafson, M. S. Hanner, A. Heck, M. Hor{\'a}nyi,
J. Kissel, B. A. Lindblad, D. Linkert, G. Linkert, I. Mann, 
J. A. M. McDonnell, G. E. Morfill, C. Polanskey, R. Riemann,
G. Schwehm, and R. Srama.
\newblock Three years of Galileo dust data: II 1993 to 1995.
\newblock to appear in {\em Planetary and Space Science}.

\bibitem[{\it Lallement and Bertin,} 1992]{lallement92}
\reference
R.~{Lallement} and P.~{Bertin}.
\newblock Northern-hemisphere observations of nearby interstellar gas -
  possible detection of the local cloud.
\newblock {\em Astronomy and Astrophysics}, 266:479--485, 1992.

\bibitem[{\it Landgraf,} 1998]{landgraf98b}
\reference
M.~Landgraf.
\newblock {\em {M}odellierung der {D}ynamik und {I}nterpretation der
  {I}n-{S}itu-{M}essung interstellaren {S}taubs in der lokalen {U}mgebung des
  {S}onnensystems}.
\newblock PhD thesis, Ruprecht--Karls--Universit{\"a}t Heidelberg,
1998.

\bibitem[{\it Landgraf and Gr\"un,} 1998]{landgraf98a}
\reference
M.~Landgraf and E.~Gr{\"u}n.
\newblock In situ measurements of interstellar dust.
\newblock In D.~Breitschwerdt, M.~J. Freyberg, and J.~Tr{\"u}mper, editors,
  {\em Proceedings of the IAU Colloquium No. 166 ``The Local Bubble and
  Beyond''}, volume 506 of {\em Lecture Notes in Physics}, pages 381--384.
  Springer Heidelberg, 1998.

\bibitem[{\it Landgraf,} 1999]{landgraf99}
\reference
M.~Landgraf.
\newblock Modeling the Motion and Distribution of Interstellar Dust
inside the Heliosphere.
\newblock {\em Journal of Geophysical Research}, this issue, 1999.

\bibitem[{\it Li and Greenberg,} 1997]{li97}
\reference
A.~{Li} and J.~M. {Greenberg}.
\newblock A unified model of interstellar dust.
\newblock {\em Astronomy and Astrophysics}, 323:566--584, 1997.

\bibitem[{\it Mathis et al.,} 1977]{mathis77}
\reference
J.~S. Mathis, W.~Rumpl, and K.~H. Nordsieck.
\newblock The size distribution of interstellar grains.
\newblock {\em Astrophysical Journal}, 280:425, 1977.

\bibitem[{\it Mathis,} 1990]{mathis90}
\reference
J.~S. Mathis.
\newblock Interstellar dust and extinction.
\newblock {\em Annual Review of Astronomy and Astrophysics}, 28:37--70, 1990.

\bibitem[{\it Morfill and Gr\"un,} 1979]{morfill79a}
\reference
G.~E. Morfill and E.~Gr{\"u}n.
\newblock The motion of charged dust particles in interplanetary space -- ii.
  interstellar grains.
\newblock {\em Planetary and Space Science}, 27:1283--1292, 1979.

\bibitem[{\it Oort and van~de~Hulst,} 1946]{oort46}
\reference
J.~H. Oort and H.~C. van~de Hulst.
\newblock {\em Bull. Astron. Inst. Netherlands}, 10:187, 1946.

\bibitem[{\it Taylor et al.,} 1996]{taylor96}
\reference
A.~Taylor, W.~J. Baggaley, and D.~I. Steel.
\newblock Discovery of interstellar dust entering the earth's atmosphere.
\newblock {\em Nature}, 380:323--325, 1996.

\bibitem[{\it Witte et al.,} 1993]{witte93}
M.~Witte, H.~Rosenbauer, H.~Banaszkiewicz, and H.~Fahr.
\newblock The ulysses neutral gas experiment - determination of the velocity
  and temperature of the interstellar neutral helium.
\newblock {\em Advances in Space Res.}, 13:(6)121--(6)130, 1993.

\end{thebibliography}
\end{document}